\documentclass[draftcls,a4paper,onecolumn,12pt]{IEEEtran}

\usepackage{multirow,multicol}
\usepackage[dvips]{color}
\usepackage{graphicx,epsfig,subfigure,pstricks}
\usepackage[noadjust]{cite}
\usepackage{amsmath,amsfonts,setspace,here}

\newcommand{\dhat}[1]{\hat{\hat{#1}}}

\newtheorem{remark}{Remark}

\ifCLASSOPTIONonecolumn
    \usepackage[nomarkers]{endfloat}
    
    \AtBeginDelayedFloats{} % For changing baseline in the figures, which are floated at the end.
\fi

\begin{document}
\title{Error Performance Analysis of DF and AF Multi-way Relay Networks with BPSK Modulation}
\author{\authorblockN{Shama N. Islam, Parastoo Sadeghi and Salman Durrani}\\
\authorblockA{Research School of Engineering\\
College of Engineering and Computer Science\\
The Australian National University, Canberra, Australia\\
E-mail: \{shama.islam, parastoo.sadeghi,
salman.durrani\}@anu.edu.au}} \maketitle

\vspace{-30pt}
\begin{abstract}
In this paper, we analyze the error performance of decode and
forward (DF) and amplify and forward (AF) multi-way relay networks
(MWRN). We consider a MWRN with pair-wise data exchange protocol
using binary phase shift keying (BPSK) modulation in both additive
white Gaussian noise (AWGN) and Rayleigh fading channels. We
quantify the possible error events in an $L$-user DF or AF MWRN and
derive accurate asymptotic bounds on the probability for the general
case that a user incorrectly decodes the messages of exactly $k$
($k\in[1,L-1]$) users. We show that at high signal-to-noise ratio
(SNR), the higher order error events ($k\geq3$) are less probable in
AF MWRN, but all error events are equally probable in a DF MWRN. We
derive the average BER of a user in a DF or AF MWRN in both AWGN and
Rayleigh fading channels under high SNR conditions. Simulation
results validate the correctness of the derived expressions. Our
results show that at medium to high SNR, DF MWRN provides better
error performance than AF MWRN in AWGN channels even with a large
number of users (for example, $L=100$). Whereas, AF MWRN outperforms
DF MWRN in Rayleigh fading channels even for much smaller number of
users (for example, $L > 10$).
\end{abstract}


\ifCLASSOPTIONonecolumn
    \newpage
\fi

\section{Introduction}\label{sec:intro}
Two-way relay networks (TWRNs), with physical-layer network coding
(PNC) protocol, have emerged as a spectrally efficient method for
bidirectional communication and exchange of information between two
nodes~\cite{Zhang-2006,Katti:2007,Rankov-2005}. In such systems, the
relay utilizes the additive nature of physical electromagnetic waves
and either amplifies and forwards (AF)~\cite{Louie-2010} or decodes
and forwards (DF)~\cite{Katti:2007,Zhang-2006} the sum of the
signals before re-transmission. Both users extract the message of
the other user by canceling self-information. Compared to other TWRN
protocols, such as digital network coding in which the relay
performs XOR operations on bit streams from the two
users~\cite{Katti1-2006,Zhang-2006}, it has been shown that PNC
requires smaller number of time slots for full information
exchange~\cite{Louie-2010}. The performance of TWRNs with PNC has
been thoroughly analyzed from the perspective of
capacity~\cite{rankov:2006,Gunduz1:2008}, bit error rate
(BER)~\cite{Louie-2010,MJu:2010,Hwang:2011,Song:2011,Cui:2008,You:2010,Zhang-2006,Zhao:2010}
and practical issues such as channel estimation and
synchronization~\cite{Lu-2012,abdallah-2012,Wang-2012,Jiang-2010}.

Recently, TWRNs have been generalized to multi-way relay networks
(MWRNs) in which multiple users can exchange information with the
help of a relay terminal~\cite{Gunduz:2009}. Potential applications
of MWRNs include file sharing in a peer-to-peer wireless network,
local measurement exchange in a sensor network or base station
information exchange in a satellite communication
network~\cite{Gunduz:2012}. Different protocols have been proposed
for MWRNs, e.g., complex field network coding which entails
symbol-level operations incorporating complex field coefficients at
the physical layer \cite{Wang:2008} and MWRNs with pair-wise data
exchange where the relay decodes or amplifies pair-wise functions of
users' messages~\cite{Ong:2010}. In particular, it was shown
in~\cite{Ong:2010} that pair-wise DF (at the relay) for binary MWRN
is theoretically the optimal strategy since it achieves the
common-rate capacity. Optimal user pairing for asymmetric MWRNS,
where users have different channel conditions, are studied
in~\cite{Noori-2012}. Practical coding schemes, based on low-density
parity-check codes, for MWRNs are proposed in~\cite{Timo-2013}.
However, a significant practical issue in MWRNs with pair-wise data
exchange is error propagation. For example in a DF MWRN, if a user
wrongly decodes another user's message, then this error propagates
through the subsequent decoding operations unless another error is
made. In an AF MWRN, the mean of the received signal is shifted from
its true value due to an earlier error. This can have a significant
impact on the average BER for a user in a MWRN.

To the best of our knowledge, an analytical characterization of the
error propagation in a MWRN has not been fully addressed in the
literature to date. The probability for the special cases that a
user incorrectly decodes the messages of exactly $k=0$, $k=1$, $k=2$
and $k = L-1$ users, respectively, in an $L$-user DF MWRN is derived
in our preliminary work in~\cite{Shama:2011}. The probability for
the special case of having at least one error event ($k\geq1$) for
AF MWRN is derived in~\cite{Amarasuriya-2012}. Apart
from~\cite{Shama:2011} and~\cite{Amarasuriya-2012}, there has been
no attempt to analyze the error performance of MWRNs with pair-wise
data exchange. There are two major limitations of these prior works.
Firstly, the derived probabilities represent certain special cases
of the more general problem of finding the probability of $k$ error
events ($k\in[1,L-1]$), i.e., where $k$ can take any integer value
in the set $[1,L-1]$. The prior works do not address the problem of
finding the probability of higher order error events ($k \ge 3$).
Secondly, the probabilities of discrete error events offer only a
partial view of the overall error performance. From the perspective
of the overall system performance, the average BER is a more useful
metric since it takes all the error events into account. The prior
works leave this as an important open problem~\cite[page
524]{Amarasuriya-2012}.

In this paper, we are concerned with the error performance analysis
of DF and AF MWRNs with BPSK Modulation. In particular, we address
the following open problems:
\begin{enumerate}
\item How can we characterize the probability of $k$ error events in DF and AF
MWRN?
\item What is the average BER for a user in a DF or AF MWRN?
\item For a given number of users and operating signal-to-noise
ratio (SNR), what is the best relaying strategy (DF or AF) in MWRN?
\end{enumerate}

As an outcome of our analysis, we obtain the following solutions to the above problems:

\begin{itemize}

 \item We derive accurate asymptotic bounds on the error probability for the general case of $k$ error events in an $L$-user DF or AF MWRN (cf. \eqref{48} and \eqref{51b}). These bounds are based on the insights gained from the analysis of the exact probability that a user incorrectly decodes the messages of $k=1$ and $k=2$ users. We show that the derived asymptotic bounds are accurate at mid to high SNR range.

\item Our analysis of the error probability for the general case of $k$ error events shows that at high SNR (a) the dominant factor in the error propagation in DF MWRN is the probability of consecutive erroneous messages resulting from a single erroneous network coded bit, (b) the dominant factor in the error propagation in AF MWRN is the probability of consecutive errors involving the middle or end users in the transmission protocol and (c) the higher order error events ($k\geq3$) are less probable in AF MWRN, but all error events are equally probable in a DF MWRN. This affects their BER sensitivity to the number of users in the system, as discussed later.

\item We use the asymptotic bounds on the probability of $k$ error events to derive closed-form expressions for the average BER of a user in DF or AF MWRN under high SNR conditions (cf. \eqref{50a} and \eqref{50b}). For both DF and AF MWRN in AWGN channel, the derived BER expressions can accurately predict the average BER of a user in medium to high SNR. For Rayleigh fading channels, the analytical expressions are within 1 dB of the simulation results at high SNR.

\item We show that for a given number of users in an AWGN channel, AF MWRN is slightly better than DF MWRN at low SNR, while DF MWRN is better than AF MWRN at medium to high SNRs. This is true even for a large number of users (for example, $L=100$). For fading channels, AF MWRN begins to outperform DF MWRN for the number of users as low as $L \approx10$. We attribute this to the lower probability of high-order error events in AF MWRN, which makes it more robust to the increase in the number of users in terms of average BER.

    \end{itemize}

The rest of the paper is organized in the following manner. The
system model is presented in Section~\ref{tools}. The challenges
associated with the characterization of the error performance in
MWRNs are discussed in Section \ref{sec:ber}. The asymptotic bounds
on the error probability for the general case of $k$ error events
and the average BER for a user in DF and AF MWRNs are derived in
Section \ref{DF} and Section \ref{AF}, respectively. The analysis is
extended to include Rayleigh fading in Section \ref{fading}. Section
\ref{number} provides the simulation results for verification of the
analytical solutions. Finally, conclusions are provided in Section
\ref{conclusion}.

Throughout this paper, we have used the following notation:
$\bigoplus$ denotes XOR operation, $\hat{(\cdot)}$ and
$\dhat{(\cdot)}$ denote decoded values at the relay and users
respectively, $\mid\cdot\mid$ denotes absolute value of a complex
variable, $\arg{(\cdot)}$ denotes the argument, $\min{(\cdot)}$
denotes the minimum value, $E[\cdot]$ denotes the expected value of
a random variable and $Q(\cdot)$ is the Gaussian Q-function.

%%%%%%%%%%%%%%%%%%%%%%%%%%%%%%%%%%%%%%%%%%%%%%%%%%%%%%%%%%%%%%%%%%%%%%%%%%%%%%%%
\section{System Model}\label{tools}

\begin{figure}
\centering
 {\subfigure[MAC phase]{
  \includegraphics[width=0.5\textwidth]{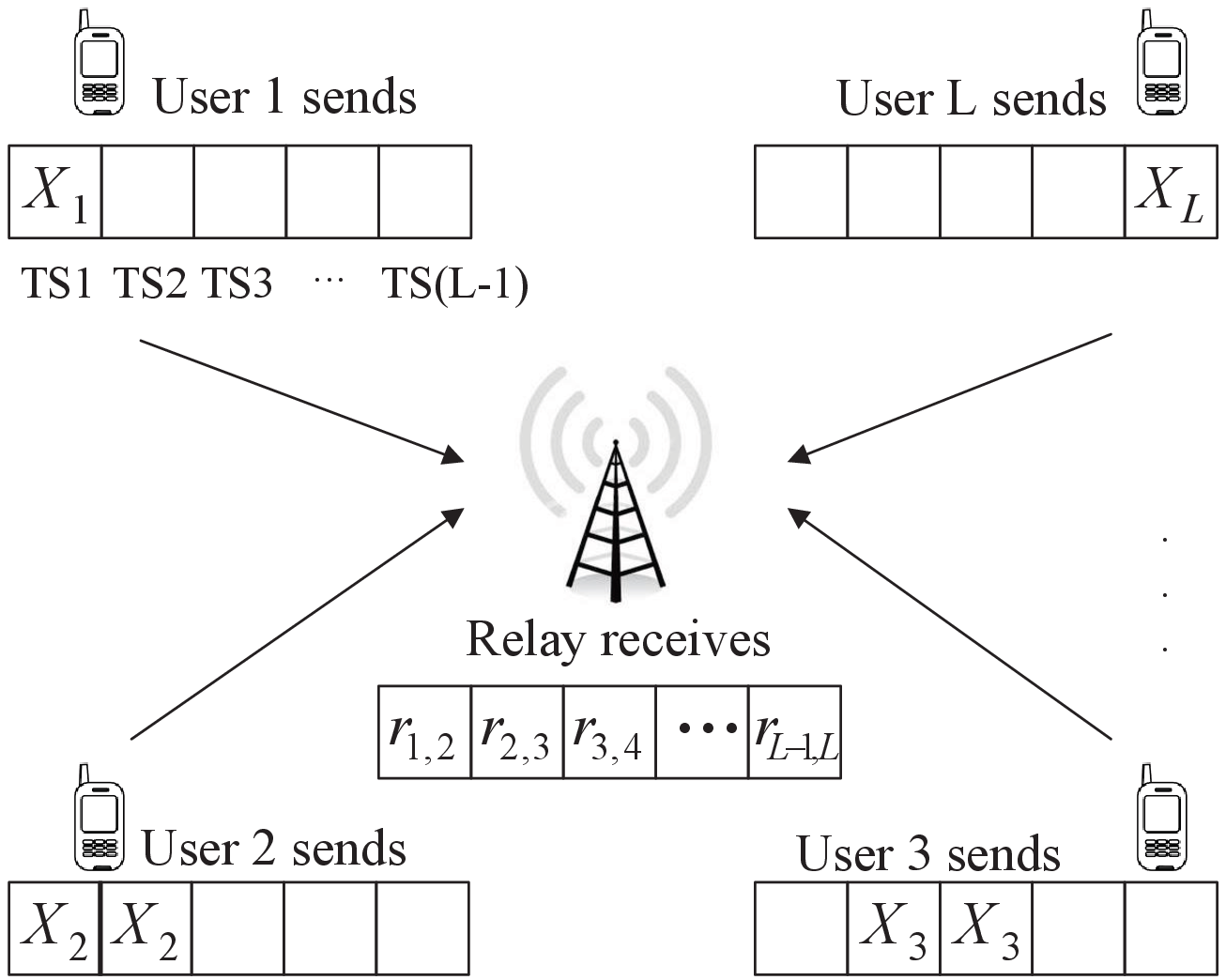}
  \label{fig:sysmodel1}
  }
\subfigure[BC phase]{
  \includegraphics[width=0.5\textwidth]{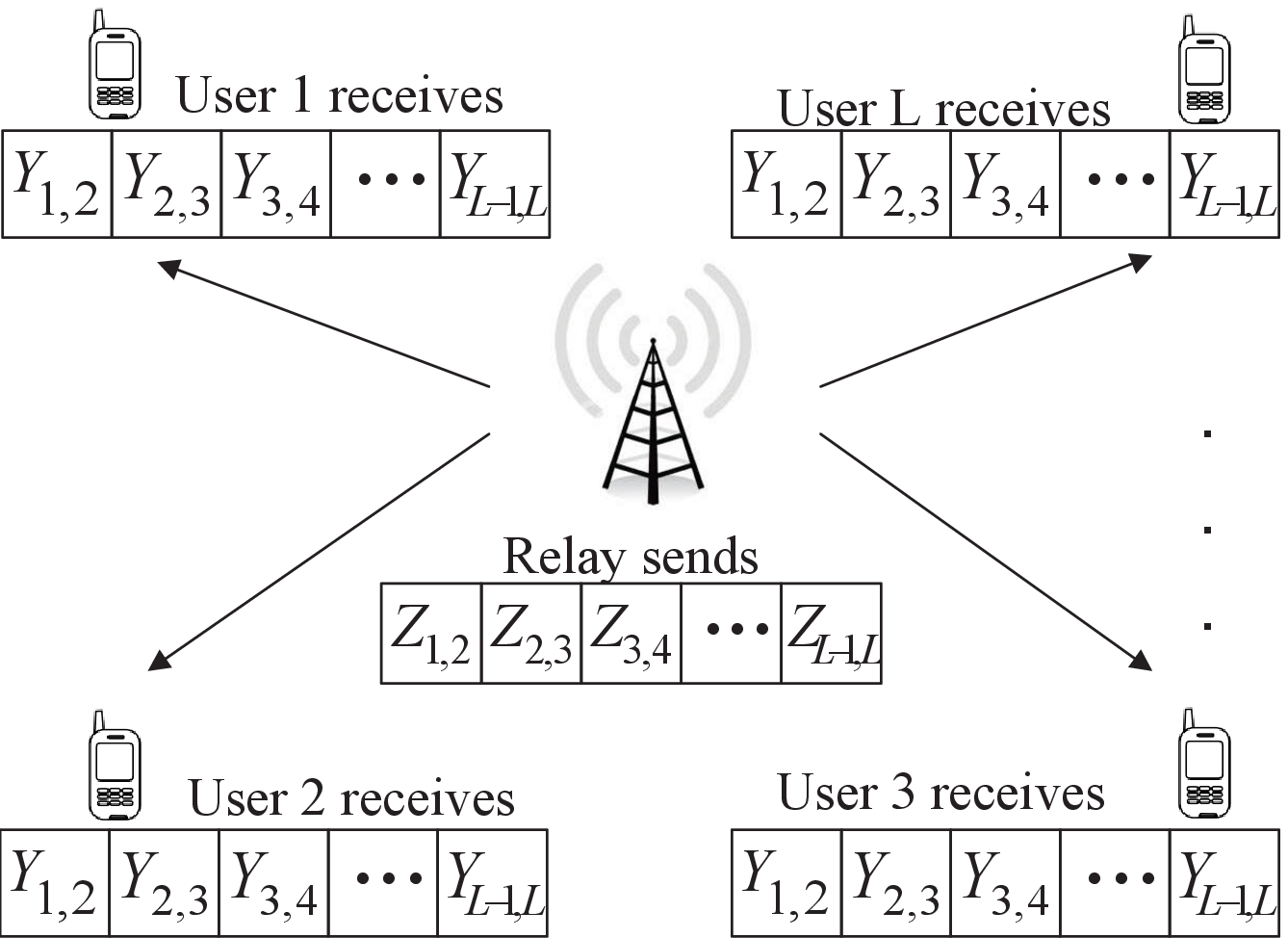}
  \label{fig:sysmodel2}
 }}
  \caption{System model for an $L$-user decode and forward (DF) multi-way relay network
(MWRN), where the users exchange information with each other via the
relay $R$. Here, TS means time slot and the other mathematical symbols are explained in Sections~\ref{sec:sysmodel1} and~\ref{sec:sysmodel2}.}
\end{figure}

Consider a multi-way relay network (MWRN) with $L$ user nodes and a
single relay node $R$. We assume that (i) there is no direct link between
the users and they exchange their information through the relay,
(ii) each node has a single antenna and operates in a half-duplex
mode, i.e., a node cannot transmit and receive simultaneously and
(iii) the MWRN operates in time-division duplex (TDD) mode, i.e.,
the uplink and downlink channels are differentiated in time slots
but occupy the same frequency slot. We concentrate on a MWRN in
which all user transmissions consist of $T$ binary phase shift
keying (BPSK) modulated symbols per frame and all the channels are
corrupted by additive white Gaussian noise (AWGN) only. Later in
Section \ref{fading}, we extend the model to Rayleigh fading
channels.

The communication among the users takes place in two phases, with
each phase comprising $L-1$ time slots \cite{Ong:2010}. In the first
\textit{multiple access phase}, users take turns to simultaneously
transmit in a pair-wise manner. Overall, the first and the last user
transmit once only while the remaining $L-2$ users transmit twice.
This phase is independent of the transmission protocol used at the
relay. In the second \textit{broadcast phase}, depending on the
relay transmission protocol, the relay broadcasts the decoded or
amplified network coded message to all the users. At the completion
of the broadcast phase, all the users have the network coded
messages corresponding to each user pair. Then they utilize self
information to extract the messages from the other users. This is
illustrated in subfigures~\ref{fig:sysmodel1}
and~\ref{fig:sysmodel2} for an $L$-user DF MWRN.

\subsection{Transmission Protocol at the Users (for both DF and AF)}\label{sec:sysmodel1}
Let the $i^{\textrm{th}}$ and $(i+1)^{\textrm{th}}$ user transmit
binary symbols, $W_{i}$ and $W_{i+1}$, which are BPSK modulated to
$X_{i}$ and $X_{i+1}$ respectively, where $W_{i} \in \{0,1\}$,
$X_{i} \in \{\pm1\}$ and $i=1,2,\hdots,L-1$. The relay receives the
signal
\begin{equation}\label{eq:1}
    r_{i,i+1}=X_{i} + X_{i+1}+n_{1},
\end{equation}

\noindent where $n_{1}$ is the zero mean AWGN in the user-relay link
with noise variance $\sigma^2_{n_1}$. For a fair comparison between
TWRNs and MWRNs later in our simulations, we maintain the same
average power per user in a MWRN as that in a TWRN and set
$\sigma^2_{n_1} = \frac{2L-2}{L}\frac{N_0}{2}$, where
$\frac{N_0}{2}$ is the noise variance in TWRN. In addition, we
assume equal power at the users and the relay, which are normalized
to one unit. Thus, the SNR per bit per user can be defined as
\begin{equation}\label{eq:snr}
    \rho=\frac{1}{\left(\frac{2L-2}{L}\right)N_0}.
\end{equation}

\noindent Depending on the relay protocol (i.e., DF or AF), the
relay makes use of the received signal $r_{i,i+1}$ in different
ways, which is discussed in the next two subsections.

\subsection{Transmission Protocol at the Relay for Decode and Forward}\label{sec:sysmodel2}
The relay first decodes the superimposed received signal
$r_{i,i+1}$ (as illustrated in Fig. \ref{fig:sysmodel2}), using the maximum a posteriori (MAP) criterion, to
obtain $\hat{V}_{i,i+1}$, which is an estimate of the true network
coded symbol, $V_{i, i+1}=W_{i}\oplus W_{i+1}$, transmitted by the
users. The optimum threshold, $\gamma_r$, for MAP detection at the
relay is derived in \cite{Zhang-2006} and is defined later in Section III
after \eqref{10}. The relay then performs BPSK modulation on the
recovered network coded symbol and retransmits to all the users,
which receive a noisy version of the signal as
\begin{equation}\label{3}
    Y_{i,i+1}=Z_{i,i+1}+n_{2},
\end{equation}

\noindent where $Z_{i,i+1} \in \{\pm1\}$ and $n_{2}$ is the zero
mean AWGN in the relay-user link with noise variance
$\sigma^2_2 = \frac{2L-2}{L}\frac{N_0}{2}$.

Each user receives and decodes the signal $Y_{i,i+1}$ (illustrated in Fig. \ref{fig:sysmodel2}) using MAP
criterion to obtain the network coded symbol $\dhat{V}_{i, i+1}$.
The optimum threshold, $\gamma$, for MAP detection at the users is
derived in \cite{Shama:2011} and is defined later in Section III
after \eqref{10}. After decoding the network coded information of
all the user pairs, the $i^{\textrm{th}}$ user performs XOR
operation between its own information symbols $W_{i}$ and the
decoded symbols $\dhat{V}_{i, i+1}$ to extract the information of
the $(i+1)^{\textrm{th}}$ user as
\begin{equation}\label{eq:df:process}
\dhat{W}_{i+1}=\dhat{V}_{i, i+1}{\oplus}W_{i}
\end{equation}
Then the $i^{\textrm{th}}$ user utilizes this extracted
information of the $(i+1)^{\textrm{th}}$ user to obtain the
information of the $(i+2)^{\textrm{th}}$ user in the same manner.
This process is continued until all the users' transmitted
information is recovered. The sequential downward information
extraction process can be expressed as
\begin{equation}\label{eq:df:process:L}
\dhat{W}_{i+2}=\dhat{V}_{i+1, i+2}{\oplus}\dhat{W}_{i+1}, \hdots,
\dhat{W}_{L}=\dhat{V}_{L-1, L}{\oplus}\dhat{W}_{L-1}
\end{equation}
Note that
for all users other than the first user, the sequential upward
information extraction process is also performed, i.e.,
$\dhat{W}_{i-1}=\dhat{V}_{i-1, i}{\oplus}W_{i}$,
$\dhat{W}_{i-2}=\dhat{V}_{i-2, i-1}{\oplus}\dhat{W}_{i-1}$,$\hdots$,
$\dhat{W}_{1}=\dhat{V}_{1, 2}{\oplus}\dhat{W}_{2}$.

\subsection{Transmission Protocol at the Relay for Amplify and Forward}
The relay amplifies the superimposed received signal $r_{i,i+1}$
with an amplification factor $\alpha$ and then retransmits to all
the users, which receive a noisy version of this retransmitted
signal as
\begin{equation}\label{7}
{Y}_{i,i+1}=\alpha(X_{i}+X_{i+1}+n_{1})+n_{2}. %
\end{equation}

\noindent where
$\alpha=\sqrt{\frac{1}{2+\frac{2L-2}{L}\frac{N_0}{2}}}$ is chosen to
maintain unity power at the users and the relay.

The $i^{\textrm{th}}$ user subtracts its own signal multiplied by
$\alpha$ from the received signal ${Y}_{i,i+1}$ and then performs
maximum likelihood (ML) detection on the resulting signal to
estimate the message of the $(i+1)^{\textrm{th}}$ user as
\begin{equation}\label{eq:AF:process}
\dhat{W}_{i+1}=\arg\min_{X_{i}\in\{\pm1\}}\mid{Y}_{i, i+1}-\alpha
X_{i}\mid^{2}
\end{equation}
Then, the $i^{\textrm{th}}$ user utilizes the BPSK modulated version of this extracted
information, i.e., $\dhat{X}_{i+1}$ to obtain the
information of the $(i+2)^{\textrm{th}}$ user in the same manner.
This process is continued until all the users' transmitted
information is recovered. The sequential downward information
extraction process can be expressed as
\begin{equation}\label{eq:AF:process:L}
\dhat{W}_{i+2}=\arg\min_{\dhat{X}_{i+1}\in\{\pm1\}}\mid{Y}_{i+1,
i+2}-\alpha\dhat{X}_{i+1}\mid^2,\hdots,
\dhat{W}_{L}=\arg\min_{\dhat{X}_{L-1}\in\{\pm1\}}\mid{Y}_{L-1,L}-\alpha\dhat{X}_{L-1}\mid^{2}.
\end{equation}
Note that for all users other than the first user, the sequential
upward information extraction process can similarly be performed.

\section{Characterizing the Error Performance in a MWRN}\label{sec:ber}
In this section, we discuss the different metrics used to
characterize the error performance in a MWRN. We also highlight the
challenges associated with calculating these metrics.

For an error-free communication, each user in a MWRN must correctly
decode the information from \textit{all} other users. Depending on
the number of users whose information is incorrectly decoded by a
certain user, different error events can occur. As highlighted
earlier in Section~\ref{sec:intro}, previous works have focused on
characterizing the special cases of error events $k=0,1,2,L-1$
\cite{Shama:2011} for DF and $k\geq1$ \cite{Amarasuriya-2012} for
AF. The error probability for the general case of $k$ error events
in an $L$-user DF or AF MWRN has not been addressed. In addition,
these discrete error events offer only a partial view of the overall
error performance. For a complete characterization of the error
performance, we need a metric that takes into account all the error
events, as well as their relative impacts. Hence, in this paper, we
also consider the average BER as the error performance metric for a
MWRN.

The average BER for the $i^{\textrm{th}}$ user in a MWRN can be defined as the expected probability of all the error events, that is,
\begin{equation}\label{37}
P_{i,avg}=\frac{1}{L-1}\sum_{k=1}^{L-1}kP_{i}(k),
\end{equation}

\noindent where $P_{i}(k)$, for $k\in[1,L-1]$, represents the
probability of exactly $k$ errors at the $i^{\textrm{th}}$ user, the
factor $k$ represents number of errors in $k^{th}$ error event and
$L-1$ denotes the number of possible error events. Note that the
average BER in~(\ref{37}) is the average across the information bits
of all the users decoded by a user.

The average BER depends on the probability of exactly $k$ error events, which is given by
\begin{equation}\label{38}
P_{i}(k)=\frac{\textrm{Number of events where $i^{\textrm{th}}$ user
incorrectly decodes messages of exactly $k$ users}}{\textrm{Packet
length, $T$}}.
\end{equation}

It is not straightforward to characterize the error probability
$P_{i}(k)$ for the general case of $k$ error events and consequently
the average BER for a user in a MWRN due to following two main
reasons. Firstly, in a DF or AF MWRN, the decision about each user
depends on the decision about previous users. For example, according
to \eqref{eq:df:process} and \eqref{eq:df:process:L} in a DF MWRN,
if an error occurs in the message extraction process, the error
propagates through to the following messages, until another error is
made. Also according to \eqref{eq:AF:process} and
\eqref{eq:AF:process:L} in an AF MWRN, the mean of the next signal
is shifted from its true value by the previous error. These
dependencies will be explained in detail in Sections \ref{DF} and
\ref{AF}, respectively. Secondly, while a TWRN has only one possible
error event, i.e, only one user's message can be incorrectly
decoded, an $L$-user MWRN consists of $(L-1)$ user pairs and so
$(L-1)$ error events are possible. This can be quite large,
depending on the number of users.

%In addition, the exact probability
%of occurrence of these error events also depends on the relaying
%protocol, i.e. DF or AF.

In the next two sections, we address these challenges and
characterize the error probability $P_{i}(k)$ for the general case
of $k$ error events and the average BER for a user in DF and AF
MWRN.

\section{Probability of $k$ error events and Average BER for a User in DF MWRN}\label{DF}
In this section, we first derive exact closed-form expressions for
the probability of $k=1$ and $k=2$ error events in an $L$-user DF
MWRN. Based on the insights provided by this analysis, we then
obtain an approximate expression for the probability of $k\geq3$
error events $P_{i}(k)$, which we use to obtain the average BER for
a user.

\begin{table}[t]
\caption{Illustration of the error cases for one and two error
events in a $10$-user DF MWRN. Here, \checkmark and $\times$
represent correct and incorrect detection, respectively.}
\label{table:t1} \centering
\begin{tabular}{|c| c| c c c c c c c c c| c|}
\hline
Error case & Decoding user & \multicolumn{9}{|c|} {Network coded message} & Error event \\
& $i$ & $\dhat{V}_{1,2}$ & $\dhat{V}_{2,3}$ & $\dhat{V}_{3,4}$ & $\dhat{V}_{4,5}$ & $\dhat{V}_{5,6}$ & $\dhat{V}_{6,7}$ & $\dhat{V}_{7,8}$ & $\dhat{V}_{8,9}$ & $\dhat{V}_{9,10}$ & \\
\hline
$A_{1}$ & $i\in\{1,L\}$ & $\times$ & $\times$ & \checkmark &\checkmark  &\checkmark &\checkmark &\checkmark  &\checkmark &\checkmark & 1\\
\hline
$B_{1}$ & $i\neq 1$ & $\times$ & $\checkmark$ & \checkmark &\checkmark  &\checkmark &\checkmark &\checkmark  &\checkmark &\checkmark & 1\\
\hline
$B_{1}$ & $i\neq L$ & $\checkmark$ & $\checkmark$ & \checkmark &\checkmark  &\checkmark &\checkmark &\checkmark  &\checkmark &$\times$ & 1\\
\hline
$C_{1}$ & $i\in\{1,L\}$ & $\times$ & $\checkmark$ & $\times$ &\checkmark  &\checkmark &\checkmark &\checkmark  &\checkmark &\checkmark & 2\\
\hline
$D_{1}$ & $i\neq 1,2$ & $\checkmark$ & $\times$ & $\checkmark$ &\checkmark  &\checkmark &\checkmark &\checkmark  &\checkmark &\checkmark & 2\\
\hline
$D_{1}$ & $i\neq L-1,L$ & $\checkmark$ & $\checkmark$ & $\checkmark$ &\checkmark  &\checkmark &\checkmark &\checkmark  &$\times$ &\checkmark & 2\\
\hline
$E_{1}$ & $i\in\{1,L\}$ & $\checkmark$ & $\times$ & $\times$ &\checkmark  &$\times$ &$\times$ &\checkmark  &\checkmark &\checkmark & 2\\
\hline
$F_{1}$ & $i\neq 1$ & $\times$ & \checkmark & \checkmark &\checkmark  &$\times$ &$\times$ &\checkmark  &\checkmark &\checkmark & 2\\
\hline
$F_{1}$ & $i\neq L$ & \checkmark & \checkmark & \checkmark &\checkmark  &$\times$ &$\times$ &\checkmark  &\checkmark &$\times$ & 2\\
\hline
$G_{1}$ & $i\neq 1,L$ & $\times$ & \checkmark & \checkmark &\checkmark  &\checkmark &$\checkmark$ &\checkmark  &\checkmark &$\times$ & 2\\
\hline
%4) (a) & $i=1$ & $\times$ & \checkmark & \checkmark &\checkmark  &\checkmark &$\checkmark$ &\checkmark  &\checkmark &\checkmark & $L-1$\\
%\hline
%4) (a) & $i=L$ & \checkmark & \checkmark & \checkmark &\checkmark  &\checkmark &$\checkmark$ &\checkmark  &\checkmark &$\times$ & $L-1$\\
%\hline
%4) (a) & $i=3$ & \checkmark & $\times$ & $\times$ &\checkmark  &\checkmark &$\checkmark$ &\checkmark  &\checkmark &\checkmark & $L-1$\\
%\hline
\end{tabular}
\end{table}

\subsection{Probability of $k=1$ Error Event}
A single error event in a DF MWRN occurs from:
\begin{itemize}
  \item error case $A_1$: two consecutive erroneous network
coded bits or,
  \item error case $B_1$: an error in the network coded
bits involving one of the end users.
\end{itemize}

\noindent For example, as illustrated in Table~\ref{table:t1}, error
case $A_1$ can occur when user $1$ wrongly decodes the message of
user $2$ by making consecutive errors in the detection of
$\dhat{V}_{1,2}$ and $\dhat{V}_{2,3}$. Similarly, error case $B_1$
can occur if there is an error in the decoding of $\dhat{V}_{1,2}$ at any user $i\neq1$ (or $\dhat{V}_{L-1,L}$ at any user $i\neq L$). Note
that the error examples shown in Table~\ref{table:t1} are not unique
and other combinations of errors are also possible.

Let $P_{A_1}$ and $P_{B_1}$ denote the probability of occurrence of
error cases $A_1$ and $B_1$, respectively. We have,
\begin{subequations}\label{23}
\begin{align}
P_{A_1}=(1-P_{DF})^{L-3}P_{DF}^{2} \label{23A1} \\
P_{B_1}=(1-P_{DF})^{L-2}P_{DF} \label{23B1}
\end{align}
\end{subequations}

\noindent where $P_{DF}$ is the probability that the network coded
message of any one user pair is incorrectly decoded, which is the
same as the average BER in a TWRN and is given by \cite{Shama:2011}
\begin{align}\label{10}
    P_{DF}=&\frac{1}{8}\left[\textrm{erfc}\left(\frac{-\gamma-1}{\sqrt{1/\rho}}\right) \left\{ \textrm{erf}\left(\frac{\gamma_{r}+2}{\sqrt{1/\rho}}\right)+\textrm{erf}\left(\frac{\gamma_{r}-2}{\sqrt{1/\rho}}\right)+
    2\textrm{erfc}\left(\frac{\gamma_{r}}{\sqrt{1/\rho}}\right)\textrm{erfc}\left(\frac{\gamma-1}{\sqrt{1/\rho}}\right)\right\}\right.\nonumber\\
    &+\left.\textrm{erfc}\left(\frac{1-\gamma}{\sqrt{1/\rho}}\right)\left\{\textrm{erfc}\left(\frac{\gamma_{r}+2}{\sqrt{1/\rho}}\right)+\textrm{erfc}\left(\frac{\gamma_{r}-2}{\sqrt{1/\rho}}\right)+
    2\textrm{erf}\left(\frac{\gamma_{r}}{\sqrt{1/\rho}}\right)\textrm{erfc}\left(\frac{\gamma+1}{\sqrt{1/\rho}}\right)\right\}\right]
\end{align}

\noindent where $\rho$ is the average SNR per bit per user defined
in~(\ref{eq:snr}), $\gamma_{r}=1+\frac{1}{4
\rho}\ln\left(1+\sqrt{1-e^{-8 \rho}}\right)$ \cite{Zhang-2006} and $ \gamma=\frac{1}{4
\rho}\ln\left(4\left(\textrm{erfc}\left(\frac{\gamma_{r}+2}{\sqrt{1/\rho}}\right)+\textrm{erfc}\left(\frac{\gamma_{r}-2}{\sqrt{1/\rho}}\right)+{2}\textrm{erfc}\left(\frac{\gamma_{r}}{\sqrt{1/\rho}}\right)\right)^{-1}-1\right)$
\cite{Shama:2011} are the optimum thresholds for MAP detection at the relay and user, respectively and
$\textrm{erf}(x)=\frac{2}{\sqrt{\pi}}\int_{0}^{x}e^{-t^{2}}dt$ and
$\textrm{erfc}(x)=\frac{2}{\sqrt{\pi}}\int_{x}^{\infty}e^{-t^{2}}dt$
are the error function and complementary error function,
respectively.

Note that in~(\ref{23A1}), the factor $P_{DF}^{2}$  represents the
probability of incorrectly decoding two consecutive erroneous
network coded bits from two user pairs while the factor
$(1-P_{DF})^{L-3}$ represents the probability that the network coded
messages of the remaining $L-3$ user pairs are correctly decoded.
Similarly in~(\ref{23B1}), the factor $P_{DF}$ represents the
probability of incorrectly decoding network coded bit involving an
end user while the factor $(1-P_{DF})^{L-2}$ represents the
probability that the network coded messages of the remaining $L-2$ user
pairs are correctly decoded. Recall that there are $L-1$ user pairs
in an $L$-user MWRN.

Using~(\ref{23}), the exact probability of one error event in a DF
MWRN can be expressed as
\begin{equation}\label{22}
    P_{i,DF}(1)=\left\{
    \begin{array}{ll}
    (L-3)P_{A_1}+2P_{B_1}&\mbox{$i\neq 1$ and $i \neq L$}\\
    (L-2)P_{A_1}+P_{B_1}&\mbox{$i=1$ or $i=L$}
    \end{array}\right.
\end{equation}

\noindent where the two cases arise from the consideration of the
two end users and the remaining users.

\begin{remark}
Equation (\ref{22}) represents the probability that a user incorrectly
decodes the message of exactly $1$ user in an $L$-user DF MWRN.
\end{remark}

\subsection{Probability of $k=2$ Error Events}

Two error events in a DF MWRN can occur from:
\begin{itemize}
  \item error case $C_1$: if two wrong network coded bits are separated by one correct
network coded bit or,
  \item error case $D_1$: if the network coded bit involving one end user is correct but
the following (or preceding) bit is incorrect or,
  \item error case $E_1$: if there are two pairs of consecutive erroneous network coded
bits or,
  \item error case $F_1$: if the network coded bit involving one end user, as well as two
other consecutive network coded bits, are incorrect or,
  \item error case $G_1$: if the network coded bits involving both the end users are
incorrect.
\end{itemize}

\noindent These error cases are illustrated in Table~\ref{table:t1}.
For example, error case $C_{1}$ can occur if user 1 incorrectly
decodes user 2 and 3's messages by making errors in detecting
$\dhat{V}_{1,2}$ and $\dhat{V}_{3,4}$. Other error cases can
similarly be explained.

Let $P_{C_1}$, $P_{D_1}$, $P_{E_1}$, $P_{F_1}$ and $P_{G_1}$ denote
the probability of occurrence of these five error cases. Using
similar logic as before, we can express these probabilities as
\begin{subequations}\label{21}
\begin{align}
    P_{C_1}&=(1-P_{DF})^{L-3}P_{DF}^{2}\\
    P_{D_1}&=(1-P_{DF})^{L-2}P_{DF}\\
    P_{E_1}&=(1-P_{DF})^{L-5}P_{DF}^{4}\\
    P_{F_1}&=(1-P_{DF})^{L-4}P_{DF}^{3}\\
    P_{G_1}&=(1-P_{DF})^{L-3}P_{DF}^{2} = P_{C_1}
\end{align}
\end{subequations}

\noindent where $P_{DF}$ is given in~(\ref{10}). Then,
using~(\ref{21}), the exact probability of two error events in a DF
MWRN can be expressed as
\begin{equation}\label{29}
    P_{i, DF}(2)=
    \left\{
    \begin{array}{ll}
    (L-3)P_{C_1}+P_{D_1}+
    \sum_{m=2}^{L-3}(L-2-m)P_{E_1}+\\
    (L-3)P_{F_1}, &\mbox{$i=1$ or $i=L$}\\
    (L-4)P_{C_1}+P_{D_1}+\sum_{m=2}^{L-4}(L-3-m)P_{E_1}+\\
    2(L-4)P_{F_1}+
    P_{C_1}, &\mbox{$i=2$ or $i=L-1$}\\
    (L-5)P_{C_1}+2P_{D_1}+
    \sum_{m=2}^{L-4}(L-3-m)P_{E_1}+\\
    2(L-4)P_{F_1}+
    P_{C_1}, &\mbox{$i=3$ or $i=L-2$}\\
    (L-5)P_{C_1}+2P_{D_1}+
    \sum_{m=2}^{i-2}(L-4-m)P_{E_1}+\\
    \sum_{m=i-1}^{L-i-1}(L-3-m)P_{E_1}+
    \sum_{m=L-i}^{L-3}(L-2-m)P_{E_1}+\\
    2(L-4)P_{F_1}+P_{C_1},&\hspace{-35pt}\mbox{$i{\notin}\left\{1, 2, 3, L-2, L-1, L\right.\}$}
    \end{array}
    \right.
\end{equation}

\noindent where $m$ is the decoding order difference between the two
users that are incorrectly decoded and $(L-3-m)$ indicates the
number of such user pairs. For example, if $i=2$, $L=10$ and $m=2$,
then user $2$ can make error about $(10-3-2)=5$ user pairs (i.e.
user pair $(3,5)$, $(4,6)$, $(5,7)$, $(6,8)$ or $(7,9)$. In this
case, messages of users $3$ and $5$ can be incorrectly decoded by
wrong detection of $\dhat{V}_{2,3}$, $\dhat{V}_{3,4}$,
$\dhat{V}_{4,5}$ and $\dhat{V}_{5,6}$.

\begin{remark}
Equation (\ref{29}) represents the probability that a user
incorrectly decodes the messages of exactly $2$ users in an $L$-user
DF MWRN.
\end{remark}

\subsection{Probability of $k$ Error Events}
The preceding subsections help to illustrate the point that finding
an exact general expression for the probability of $k$ error events,
where $k\geq3$, is difficult due to the many different ways $k$
error events can occur. Hence, in this subsection, we focus on
finding an approximate expression for the probability of $k$ error
events using high SNR assumption. This will be useful in deriving
the average BER in the next subsection. Note that the use of the
high SNR assumption to facilitate closed-form results is commonly
used in two-way~\cite{Louie-2010,Song:2011,Wang-2012} and other
types of relay networks~\cite{Zhao:2006,Su:2008}.

Comparing \eqref{23} and \eqref{21}, we can see that
$P_{C_1}=P_{A_1}$ and $P_{D_1}=P_{B_1}$. At high SNR, the higher
order terms involving $P_{DF}^2$ and higher powers can be neglected
and only the terms $P_{B_1}$ and $P_{D_1}$ effectively contribute to
the probability of one and two error events in \eqref{22} and
\eqref{29}, respectively. Recall that $P_{B_1}$ is the probability
of one error about the network coded message of an end user and
$P_{D_1}$ is the probability of one erroneous network coded bit
involving users just following (or preceding) the end user.
Extending this analogy to the case of $k$ error events, the
dominating factor at high SNR would represent the scenario when the
network coded bit involving the $k^{\textrm{th}}$ and
$(k+1)^{\textrm{th}}$ (or $(L-k+1)^{\textrm{th}}$ and
$(L-k)^{\textrm{th}}$) users is incorrectly decoded, resulting in
error about $k$ users' messages. Thus, the probability of $k$ error
events can be asymptotically approximated as
\begin{equation}\label{48}
P_{i,DF}(k)\approx(1-P_{DF})^{L-2}P_{DF}\approx P_{DF}
\end{equation}

\noindent where in the last step we have used the fact that at high
SNR $P_{DF} \ll 1$ and hence $(1-P_{DF})\approx1$. It will be shown in Section~\ref{number} that for medium to high SNRs, (\ref{48}) can accurately predict the probability of $k$ error events in a DF MWRN.

\begin{remark}
Equation (\ref{48}) shows that at high SNR in an $L$-user DF MWRN,
all the error events are equally probable and their probability can
be asymptotically approximated as $P_{DF}$, in \eqref{10}, i.e., the
average BER in a TWRN.
\end{remark}

\subsection{Average BER}
Substituting \eqref{48} in \eqref{37} and simplifying, the average
BER for a user in DF MWRN is
\begin{align}\label{50a}
P_{i,avg,DF}&=\left(\sum_{k=1}^{L-1}k\right)\frac{P_{DF}}{L-1}=\frac{L(L-1)}{2}\frac{P_{DF}}{L-1}\nonumber\\
    &=\frac{L}{2}P_{DF}
\end{align}

\begin{remark}
Equation \eqref{50a} shows that at high SNR, the average BER in an
$L$-user DF MWRN can be asymptotically approximated as the average
BER in a TWRN scaled by a factor of $L/2$. Although \eqref{50a} is
obtained using a high SNR assumption, it will be shown later in
Section~\ref{number} that the average BER is well approximated even
at medium to high SNRs.
\end{remark}

\section{Probability of $k$ error events and Average BER for a User in AF MWRN}\label{AF}
In this section, we characterize the average BER for a user in an
$L$-user AF MWRN. The general approach in our analysis is similar to
the case of DF MWRN, with some important differences which are
highlighted in the following subsections.

\subsection{Probability of $k=1$ Error Event}
A single error event in an AF MWRN occurs from:
\begin{itemize}
  \item error case $A_2$: a middle user's message is wrongly estimated with correct decision about the
following user or,
  \item error case $B_2$: an error in the estimated
signal of one of the end users.
\end{itemize}
\begin{table}[t]
\caption{Illustration of the error cases for one and two error
events in a $10$-user AF MWRN. Here, \checkmark and $\times$
represent correct and incorrect detection, respectively.}
\label{table:t2} \centering
\begin{tabular}{|c| c| c c c c c c c c c c| c|}
\hline
Error case & Decoding user & \multicolumn{10}{|c|} {Extracted messages} & Error event \\
& $i$ & $\dhat{X}_{1}$ & $\dhat{X}_{2}$ & $\dhat{X}_{3}$ & $\dhat{X}_{4}$ & $\dhat{X}_{5}$ & $\dhat{X}_{6}$ & $\dhat{X}_{7}$ & $\dhat{X}_{8}$ & $\dhat{X}_{9}$ & $\dhat{X}_{10}$ & \\
\hline
$A_{2}$ & $i\in\{1,L\}$ & \checkmark & $\times$ & $\checkmark$ &\checkmark  &\checkmark &\checkmark &\checkmark  &\checkmark &\checkmark & \checkmark & 1\\
\hline
$B_{2}$ & $i\neq 1$ & $\times$ & $\checkmark$ & \checkmark &\checkmark  &\checkmark &\checkmark &\checkmark  &\checkmark &\checkmark & \checkmark & 1\\
\hline
$B_{2}$ & $i\neq L$ & $\checkmark$ & $\checkmark$ & \checkmark &\checkmark  &\checkmark &\checkmark &\checkmark  &\checkmark &\checkmark & $\times$ & 1\\
\hline
$C_{2}$ & $i\in\{1,L\}$ & \checkmark & $\checkmark$ & $\times$ &$\times$  &\checkmark &\checkmark &\checkmark  &\checkmark &\checkmark & \checkmark & 2\\
\hline
$D_{2}$ & $i\neq 1,2$ & $\times$ & $\times$ & $\checkmark$ &\checkmark  &\checkmark &\checkmark &\checkmark  &\checkmark &\checkmark & \checkmark & 2\\
\hline
$D_{2}$ & $i\neq L-1,L$ & $\checkmark$ & $\checkmark$ & $\checkmark$ &\checkmark  &\checkmark &\checkmark &\checkmark  &\checkmark &$\times$ & $\times$ & 2\\
\hline
$E_{2}$ & $i\in\{1,L\}$ & $\checkmark$ & $\times$ & \checkmark &\checkmark  &$\times$ &\checkmark &\checkmark  &\checkmark &\checkmark & \checkmark & 2\\
\hline
$F_{2}$ & $i\neq 1$ & $\times$ & \checkmark & \checkmark &\checkmark  &$\times$ &\checkmark &\checkmark  &\checkmark &\checkmark & \checkmark & 2\\
\hline
$F_{2}$ & $i\neq L$ & \checkmark & \checkmark & \checkmark &\checkmark  &$\times$ &\checkmark&\checkmark  &\checkmark &\checkmark& $\times$ & 2\\
\hline
$G_{2}$ & $i\neq 1,L$ & $\times$ & \checkmark & \checkmark &\checkmark  &\checkmark &$\checkmark$ &\checkmark  &\checkmark &\checkmark & $\times$ & 2\\
\hline
%4) (a) & $i=1$ & $-$ & $\times$ & $\times$ &$\times$ &$\times$&$\times$&$\times$  &$\times$&$\times$& $\times$ &$L-1$\\
%\hline
%4) (a) & $i=L$ & $\times$ & $\times$ & $\times$ &$\times$ &$\times$ &$\times$&$\times$&$\times$&$\times$ & $-$ & $L-1$\\
%\hline
%4) (a) & $i=3$ & $\times$ & $\times$ & $-$ &$\times$  &$\times$ &$\times$ &$\times$  &$\times$ &$\times$ & $\times$ & $L-1$\\
%\hline
\end{tabular}
\end{table}

\noindent These error cases are illustrated in Table~\ref{table:t2}.

Let $P_{A_2}$ and $P_{B_2}$ denote the probability of occurrence of
error cases $A_2$ and $B_2$, respectively. We have,
\begin{subequations}\label{26}
\begin{align}
P_{A_2}&=(1-P_{AF})^{L-3}P_{AF}(1-P'_{AF}) \label{26A2} \\
P_{B_2}&=(1-P_{AF})^{L-2}P_{AF} \label{26B2}
\end{align}
\end{subequations}

\noindent where $P_{AF}$ is the probability that the message of any
one user is incorrectly decoded, which is the same as the average
BER in an AF TWRN and is given by~\cite{Cui:2008}
\begin{equation}\label{15}
P_{AF}=\frac{1}{2}\textrm{erfc}\left(\frac{\alpha}{\sqrt{\left(\alpha^{2}+1\right)(1/\rho)}}\right)
\end{equation}

\noindent where $\rho$ is the average SNR per bit per user defined
in~(\ref{eq:snr}), $\alpha$ is the amplification factor defined
below \eqref{7} and $P'_{AF}$ is the probability of wrongly
detecting the message of a user given that the previous user's
message is also incorrect. This can be easily found as follows. To
find $P'_{AF}$, we need to find the probability
$P(\dhat{W}_{i+2}\neq W_{i+2}|\dhat{W}_{i+1}\neq W_{i+1})$. If
$\dhat{X}_{i+1}\neq X_{i+1}$, then $\dhat{X}_{i+2}=\alpha
X_{i+1}+\alpha X_{i+2}+\alpha n_{1}+n_{2}-\alpha
\dhat{X}_{i+1}=\alpha X_{i+2}+\alpha n_{1}+n_{2}+2\alpha X_{i+1}$.
Thus, the mean of the received signal is shifted by either $2\alpha$
or $-2\alpha$. Using this fact and~(\ref{15}), we have
\begin{equation}\label{18}
P'_{AF}=\frac{1}{4}\left[\textrm{erfc}\left(\frac{3\alpha}{\sqrt{\left(\alpha^{2}+1\right)(1/\rho)}}\right)+
\textrm{erfc}\left(\frac{-\alpha}{\sqrt{\left(\alpha^{2}+1\right)(1/\rho)}}\right)\right]
\end{equation}

Finally, using~(\ref{26}), the exact probability of one error event
in an AF MWRN can be expressed as
\begin{equation}\label{eq:266}
    P_{i,AF}(1)=\left\{
    \begin{array}{ll}
    (L-3)P_{A_2}+2P_{B_2}&\mbox{$i\neq 1$ and $i \neq L$}\\
    (L-2)P_{A_2}+P_{B_2}&\mbox{$i=1$ or $i=L$}
    \end{array}\right.
\end{equation}

\noindent where the two cases arise from the consideration of the
two end users and the remaining users.

\begin{remark}
Assume that $X_{i+2} = 1$. While the shift of the mean of the signal by $2\alpha$ (when $X_{i+1} = 1$) is helpful in reducing the probability of error in detecting $X_{i+2} = 1$, the shift in the mean by $-2\alpha$ (when $X_{i+1} = -1$) would be seriously detrimental for its detection. We will use this fact later in our high SNR BER analysis by setting $P'_{AF}\approx \frac{1}{2}$.
\end{remark}

\begin{remark}
Equation (\ref{eq:266}) represents the probability that a user incorrectly
decodes the message of exactly $1$ user in an $L$-user AF MWRN.
(\ref{eq:266}) is different from (\ref{22}) due to the presence of~(\ref{18}), which is large even at moderate to high SNRs.
\end{remark}

\subsection{Probability of $k=2$ Error Events}

Two error events in an AF MWRN can occur from:
\begin{itemize}
  \item error case $C_2$: if messages of two consecutive users are incorrectly decoded but
the message of the user next to them is correct or,
  \item error case $D_2$: if the estimated message of the end user and that of the
following (or preceding) user are incorrect or,
  \item error case $E_2$: if two middle users' messages are incorrectly
estimated provided that the message of the users adjacent to each of
them are correct or,
  \item error case $F_2$: if there is error about the message of one end
user and any other user, provided that the messages of the users in
between them are correctly estimated or,
  \item error case $G_2$: if both the end users' messages are
incorrectly estimated.
\end{itemize}

\noindent These error cases are illustrated in Table~\ref{table:t2}.

Let $P_{C_2}$, $P_{D_2}$, $P_{E_2}$, $P_{F_2}$ and $P_{G_2}$ denote
the probability of occurrence of these five error cases. Using
similar logic as before, we can express these probabilities as
\begin{subequations}\label{24}
\begin{align}
    P_{C_2}&=(1-P_{AF})^{L-4}P_{AF}(1-P'_{AF})P'_{AF}\\
    P_{D_2}&=(1-P_{AF})^{L-3}P_{AF}P'_{AF}\\
    P_{E_2}&=(1-P_{AF})^{L-5}P^{2}_{AF}(1-P'_{AF})^{2}\\
    P_{F_2}&=(1-P_{AF})^{L-4}P^{2}_{AF}(1-P'_{AF})\\
    P_{G_2}&=(1-P_{AF})^{L-3}P^{2}_{AF} \neq P_{C_2}
\end{align}
\end{subequations}

\noindent where $P_{AF}$ and $P'_{AF}$ are given in~(\ref{15})
and~(\ref{18}) respectively. Note that the expressions for error
cases $C_2$ to $E_2$ are different from the error cases $C_1$ to
$E_1$. This is due to the different relay processing in AF and DF
MWRNs.

Using~(\ref{24}), the exact probability of two error events in an AF
MWRN can be expressed as
\begin{equation}\label{31}
    P_{i, AF}(2)=
    \left\{
    \begin{array}{ll}
    (L-3)P_{C_2}+P_{D_2}+
    \sum_{m=2}^{L-3}(L-2-m)P_{E_2}+\\
    (L-3)P_{F_2}, &\mbox{$i=1$ or $i=L$}\\
    (L-4)P_{C_2}+P_{D_2}+\sum_{m=2}^{L-4}(L-3-m)P_{E_2}+\\
    2(L-4)P_{F_2}+P_{G_2}, &\mbox{$i=2$ or $i=L-1$}\\
    (L-5)P_{C_2}+2P_{D_2}+
    \sum_{m=2}^{L-4}(L-3-m)P_{E_2}+\\2(L-4)P_{F_2}+P_{G_2}, &\mbox{$i=3$ or $i=L-2$}\\
    (L-5)P_{C_2}+2P_{D_2}+
    \sum_{m=2}^{i-2}(L-4-m)P_{E_2}+\\
    \sum_{m=i-1}^{L-i-1}(L-3-m)P_{E_2}+
    \sum_{m=L-i}^{L-3}(L-2-m)P_{E_2}+\\
    2(L-4)P_{F_2}+P_{G_2}, &\hspace{-35pt}\mbox{$i{\notin}\left\{1, 2, 3, L-2, L-1, L\right.\}$}
    \end{array}
    \right.
\end{equation}

\noindent where $m$ is the decoding order difference between the two
users that are incorrectly decoded.

\begin{remark}
Equation (\ref{31}) represents the probability that a user incorrectly
decodes the message of exactly $2$ users in an $L$-user AF MWRN.
\end{remark}

\subsection{Probability of $k$ Error Events}
As for the case of DF MWRN, it is very hard to find an exact general
expression for the probability of $k$ error events in AF MWRN.
Hence, in this subsection, we focus on finding an approximate
expression for the probability of $k$ error events using high SNR
assumption.

At high SNR, we can neglect $P_{E_{2}}$, $P_{F_{2}}$ and $P_{G_{2}}$
in \eqref{24} since they involve higher order product terms of
probabilities. Comparing \eqref{26} and \eqref{24}, we can see that
the relationship between the dominating terms in the probability of
one and two error events at high SNR is
$C_{2}=\frac{P'_{AF}}{1-P_{AF}}A_{2}$,
$D_{2}=\frac{P'_{AF}}{1-P_{AF}}B_{2}$. Recall that $C_2$ and $D_2$
correspond to the cases of two consecutive errors involving middle
users and two consecutive errors involving one of the end users,
respectively. Extending this analogy to the case of $k$ error
events, the dominating terms at high SNR would represent the cases
of $k$ consecutive errors in the middle users and $k$ consecutive
errors involving one end user and $k-1$ following (or preceding)
users. Thus, the probability of $k$ error events can be
asymptotically approximated as
\begin{align}
P_{i,AF}(k)&\approx\left(\frac{P'_{AF}}{1-P_{AF}}\right)^{k-1}\left
\{(L-k-1)(1-P_{AF})^{L-3}P_{AF}(1-P'_{AF})+(1-P_{AF})^{L-2}P_{AF}\right\}\label{51a}\\
&\approx \frac{L-k+1}{2^{k}}P_{AF}\label{51b}
\end{align}

\noindent where in the last step we have used the fact that at high
SNR $P'_{AF}\approx\frac{1}{2}$ and $1-P_{AF}\approx1$. It will be shown in Section~\ref{number} that for medium to high SNRs, (\ref{51b}) can accurately predict the probability of $k$ error events in an AF MWRN.

\begin{remark}
Equation (\ref{51b}) shows that at high SNR the probability of $k$
error events in an AF MWRN can be asymptotically approximated as the
average BER of an AF TWRN scaled by a factor $(L-k+1)/2^{k}$, which
depends on both $L$ and $k$. Comparing (\ref{51b}) and (\ref{48}) we
can see that, at high SNR, the higher order error events are less
probable in an $L$-user AF MWRN, but all error events are equally
probable in an $L$-user DF MWRN.
\end{remark}

\subsection{Average BER}
Substituting \eqref{51b} in \eqref{37} and simplifying, the average
BER for a user in AF MWRN is
\begin{align}\label{50b}
P_{i,avg,AF}&=P_{AF}\sum_{k=1}^{L-1}\frac{L-k+1}{2^{k}}\nonumber\\
    &=\left(\frac{L+1}{L-1}\left(2-\frac{L}{2^{L-2}}\right)-\frac{3}{L-1}\left(2-\frac{L^2-3}{2^{L-2}}\right)\right)P_{AF}
\end{align}

\begin{remark}
Equation \eqref{50b} shows that at high SNR the average BER in an
$L$-user AF MWRN can be asymptotically approximated as the average
BER in a MWRN scaled by a factor
$\left(\frac{L+1}{L-1}(2-\frac{L}{2^{L-2}})-\frac{3}{L-1}(2-\frac{L^2-3}{2^{L-2}})\right)$.
Comparing \eqref{50a} and \eqref{50b}, we can see that the larger
number of error events have a smaller contribution in the average
BER for a user in AF MWRN, whereas they have the same contribution
as the small number of error events in a DF MWRN.
\end{remark}

%%%%%%%%%%%%%%%%%%%%%%%%%%%%%%%%%%%%%%%%%%%%%%%%%%%%%%%%%%%%%%%%%%%%%%%%%%%%%%%%
\section{Average BER for a user in MWRN with Rayleigh
Fading}\label{fading}

In this section, we demonstrate that the preceding analysis is also
applicable for the case of DF or AF MWRN with Rayleigh fading
channels. Following~\cite{MJu:2010}, we assume that (i) all the
channels are reciprocal, which is typical in TDD systems, (ii) the
channel coefficients are modeled as independent zero-mean and
unit-variance complex-valued Gaussian random variables, (iii) the
channel coefficients are independent during the multiple access and
broadcast phases and (iv) perfect channel state information is
available at the relay and the user nodes. In practice, accurate channel state information can be obtained by sending pilot symbols \cite{MJu:2010,Amarasuriya-2012,Cui:2009}, the consideration of which is outside the scope of this paper. Taking Rayleigh fading
into account, \eqref{eq:1} modifies to
\begin{equation}\label{40}
r_{i,i+1}=h_{i}X_{i}+h_{i+1}X_{i+1}+n_{1}
\end{equation}
where $h_{i}$ and $h_{i+1}$ are the complex channel coefficients for
the $i^{\textrm{th}}$ and $(i+1)^{\textrm{th}}$ user, respectively.

\subsection{DF MWRN with Rayleigh Fading}
The relay decodes the received signal using ML criterion~\cite{MJu:2010}
and obtains an estimate of the corresponding network coded message.
The relay then broadcasts the estimated signal. Thus, \eqref{3}
modifies to
\begin{equation}\label{42}
    Y_{i,i+1}=h_{i}Z_{i,i+1}+n_{2}
\end{equation}

\noindent The users then detect the received signal through ML
criterion~\cite{MJu:2010}.

With the modified signal model, the error propagation in DF MWRN is
almost similar to the case as before. Thus, it can be shown that the
probability of large number of errors is asymptotically the same as
that of small number of errors, even in the presence of fading.
Hence, we can use \eqref{50a} to find the average BER for a user. In
order to do this, we need an expression for the BER in a DF TWRN, $P_{DF}$. No exact
expression is available in the literature for the average BER in a
TWRN with Rayleigh fading. However, upper and lower bounds have been
derived in~\cite{MJu:2010}. In this work, we use the upper bound for
$P_{DF}$, which is given by \cite{MJu:2010}
\begin{align}\label{47}
P_{DF}&=2\Phi_{1}(\bar{\gamma})+\frac{1}{2}\Xi(\bar{\gamma})
\end{align}

\noindent where $\bar{\gamma}=\rho$,
$\Phi_{1}(\bar{\gamma})=\frac{1-\sqrt{\frac{\bar{\gamma}}{1+\bar{\gamma}}}}{2}$,
$\Xi(\bar{\gamma})=2\Phi_{1}(\bar{\gamma})-4\{\Phi_{1}(\bar{\gamma})\}^{2}-2\Phi_{2}(\bar{\gamma})-2\sqrt{\frac{\bar{\gamma}}{1+\bar{\gamma}}}\Phi_{3}(\bar{\gamma})$,
$\Phi_{2}(\bar{\gamma})=\frac{1}{2\pi}[\frac{\pi}{2}-2\sqrt{\frac{\bar{\gamma}}{1+\bar{\gamma}}}(\frac{\pi}{2}-\tan^{-1}\sqrt{\frac{\bar{\gamma}}{1+\bar{\gamma}}})]$,
$\Phi_{3}(\bar{\gamma})=\frac{1}{2\pi}[\frac{\pi}{2}-\delta_{1}(\frac{\pi}{2}+\tan^{-1}\zeta_{1})-\delta_{2}(\frac{\pi}{2}+\tan^{-1}\zeta_{2})]$,
$\delta_{1}=\sqrt{\frac{1+\bar{\gamma}}{3+\bar{\gamma}}},
\delta_{2}=\sqrt{\frac{\bar{\gamma}}{2+\bar{\gamma}}}$ and
$\zeta_{j}=-\delta_{j}\cot(\sqrt{\frac{\bar{\gamma}}{1+\bar{\gamma}}})$
for $j=1,2$.

\subsection{AF MWRN with Rayleigh Fading}
For AF MWRN, the amplified and retransmitted signal in \eqref{7}
modifies to
\begin{equation}\label{44}
    Y_{i,i+1}=h_{i}\alpha(h_{i}X_{i}+h_{i+1}X_{i+1}+n_{1})+n_{2}
  \end{equation}

\noindent After subtracting self information, user $i$ performs ML
detection to estimate the other user's message. The sequential
downward and upward message extraction process is the same as
before.

With the modified signal model, the error propagation in AF MWRN is
different from the AWGN case. This is because the
primary cause of error propagation in AF MWRN is the shifting of the
mean of the received signal when the previous message has been
incorrectly detected. For example, if $\dhat{X}_{i+1}\neq X_{i+1},$
then $\dhat{X}_{i+2}=\alpha h_{i}h_{i+2}X_{i+2}+\alpha
h_{i}n_{1}+n_{2}+2\alpha h_{i}h_{i+1}X_{i+1}$. Thus we can see that
the mean of the received signal is affected by the channel
coefficients. That is why, we cannot ignore $P'_{AF}$ and obtain \eqref{51b}
from \eqref{51a}. So, instead of \eqref{51b}, we will use
\eqref{51a} to provide the analytical expression of average BER for
a user, where the exact average BER for an AF TWRN in Rayleigh
fading is given by~\cite{Cui:2009,Hwang:2011}
\begin{equation}\label{53}
P_{AF}=Q\left(\sqrt{\frac{\mid h_{i}\mid^{2}\mid
h_{i+1}\mid^{2}}{2\mid h_{i}\mid^{2}(1/\rho)+\mid
h_{i+1}\mid^{2}(1/\rho)+(1/\rho)^2}}\right)
\end{equation}

\noindent and the expression for $P'_{AF}$ is similarly derived as
\begin{equation}\label{60}
P'_{AF}=Q\left(\sqrt{\frac{\mid h_{i}\mid^{2}\mid
h_{i+2}\mid^{2}}{4\mid h_{i}\mid^{2}\mid
h_{i+1}\mid^{2}+2\mid h_{i}\mid^{2}(1/\rho)+\mid
h_{i+2}\mid^{2}(1/\rho)+(1/\rho)^2}}\right)
\end{equation}

\noindent where $Q(x)=\frac{1}{\sqrt{2\pi}}\int_{x}^\infty e^{-\frac{t^2}{2}}dt$ is the Gaussian Q-function.
%%%%%%%%%%%%%%%%%%%%%%%%%%%%%%%%%%%%%%%%%%%%%%%%%%%%%%%%%%%%%%%%%%%%%%%%%%%%%%%%
\section{Results} \label{number}

In this section, we compare the BER expressions obtained by our
analysis with the BER results obtained by Monte Carlo simulations.
We consider three cases $L=10$, $L=50$ and $L=100$ users and each
user transmits a packet of $T=10000$ bits. The SNR is assumed to be
SNR per bit per user and user 1 is assumed to be decoding the
messages of all other users. The simulation results are averaged
over $1000$ Monte Carlo trials per SNR point.

\subsection{Probability of different error events in an AWGN DF MWRN}
\begin{figure}
  \includegraphics[width=0.85\textwidth]{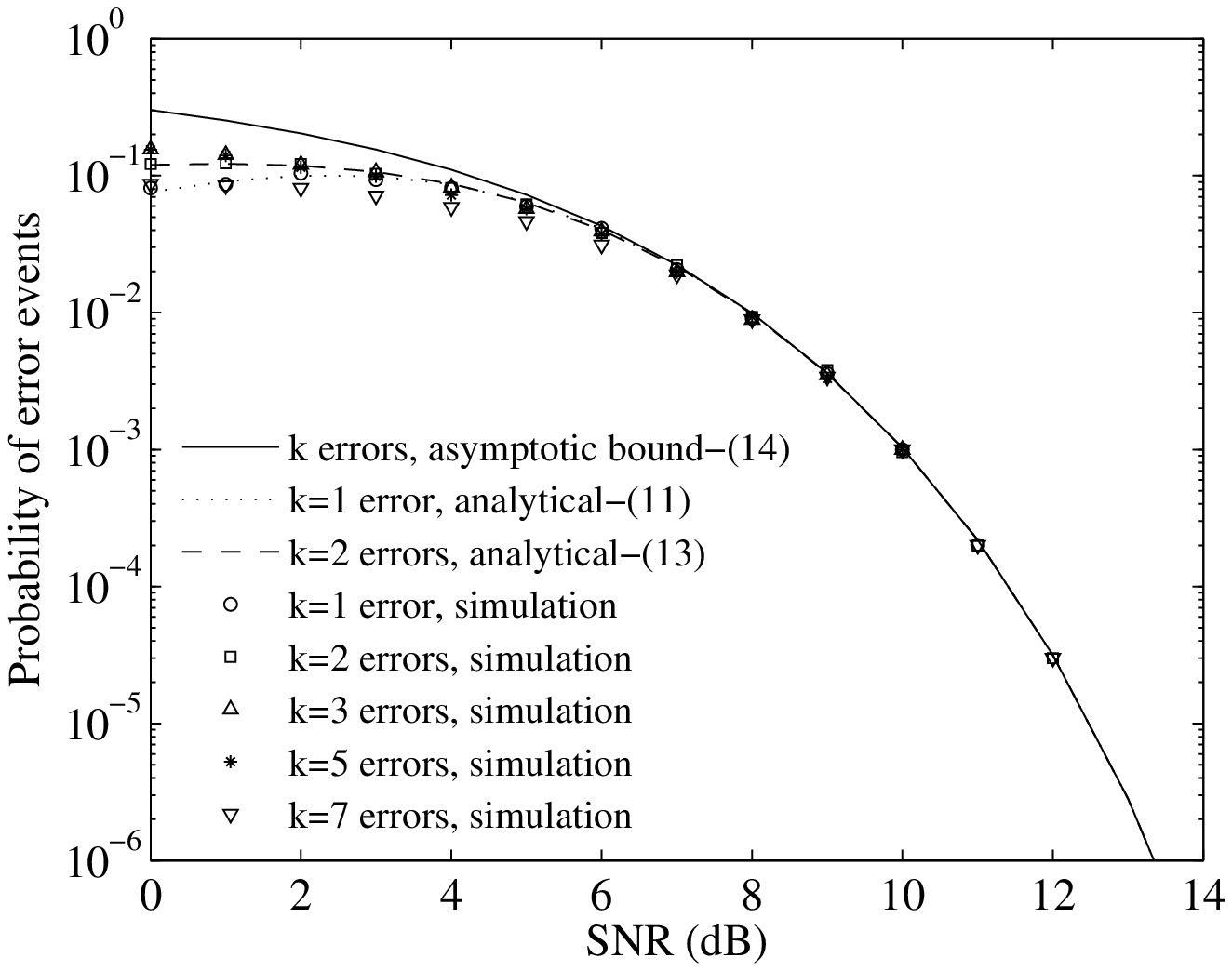}
  \caption{Probability of $k=1,2,3,5,7$ error events in an $L=10$ user DF MWRN with AWGN.}
\label{fig:Fig2}
\end{figure}

Fig.~\ref{fig:Fig2} plots the probability of $k$ error events $P_{i,DF}(k)$ in an $L=10$ user DF MWRN in the case of AWGN. The simulation results
are plotted for $k=1,2,3,5,7$ and compared with the asymptotic bound in \eqref{48}. For $k=1,2$ the exact probabilities are also plotted using~\eqref{22} and \eqref{29}, respectively. As highlighted in Remark 3, in an $L$-user DF MWRN, all the error events are equally probable and their probability can be asymptotically approximated as \eqref{48}. This is confirmed by the results in Fig.~\ref{fig:Fig2}. We can see that for medium to high SNRs (SNR $> 5$ dB), the asymptotic expression in \eqref{48} is very accurate in predicting the probability of $k$ error events, for all the considered values of $k$. This verifies the accuracy of~\eqref{48}.

\subsection{Probability of different error events in an AWGN AF MWRN}
\begin{figure}
{\subfigure[$k=1,2$]{
  \includegraphics[width=0.5\textwidth]{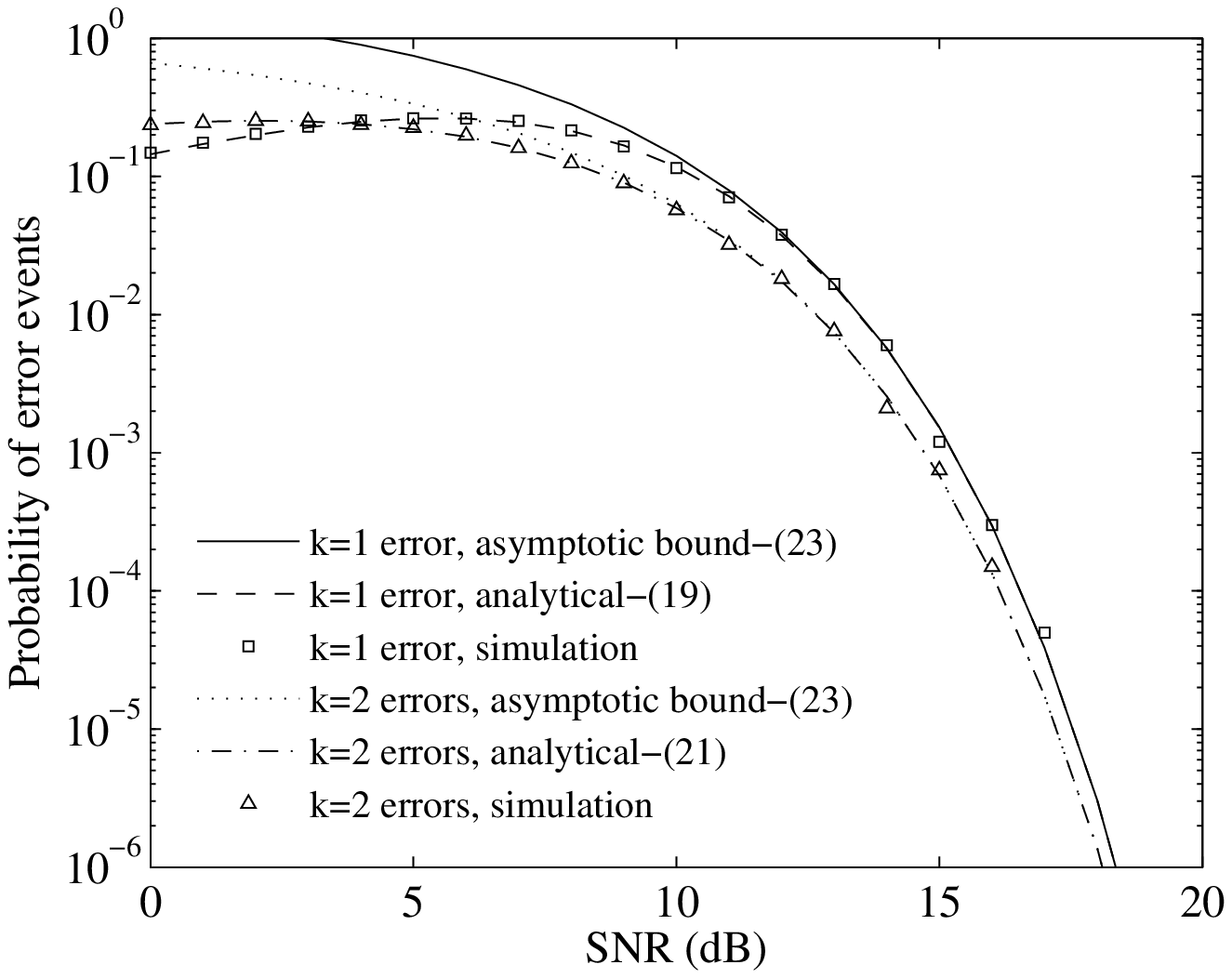}
  \label{fig:su3}}
\subfigure[$k=3,5,7$]{
  \includegraphics[width=0.5\textwidth]{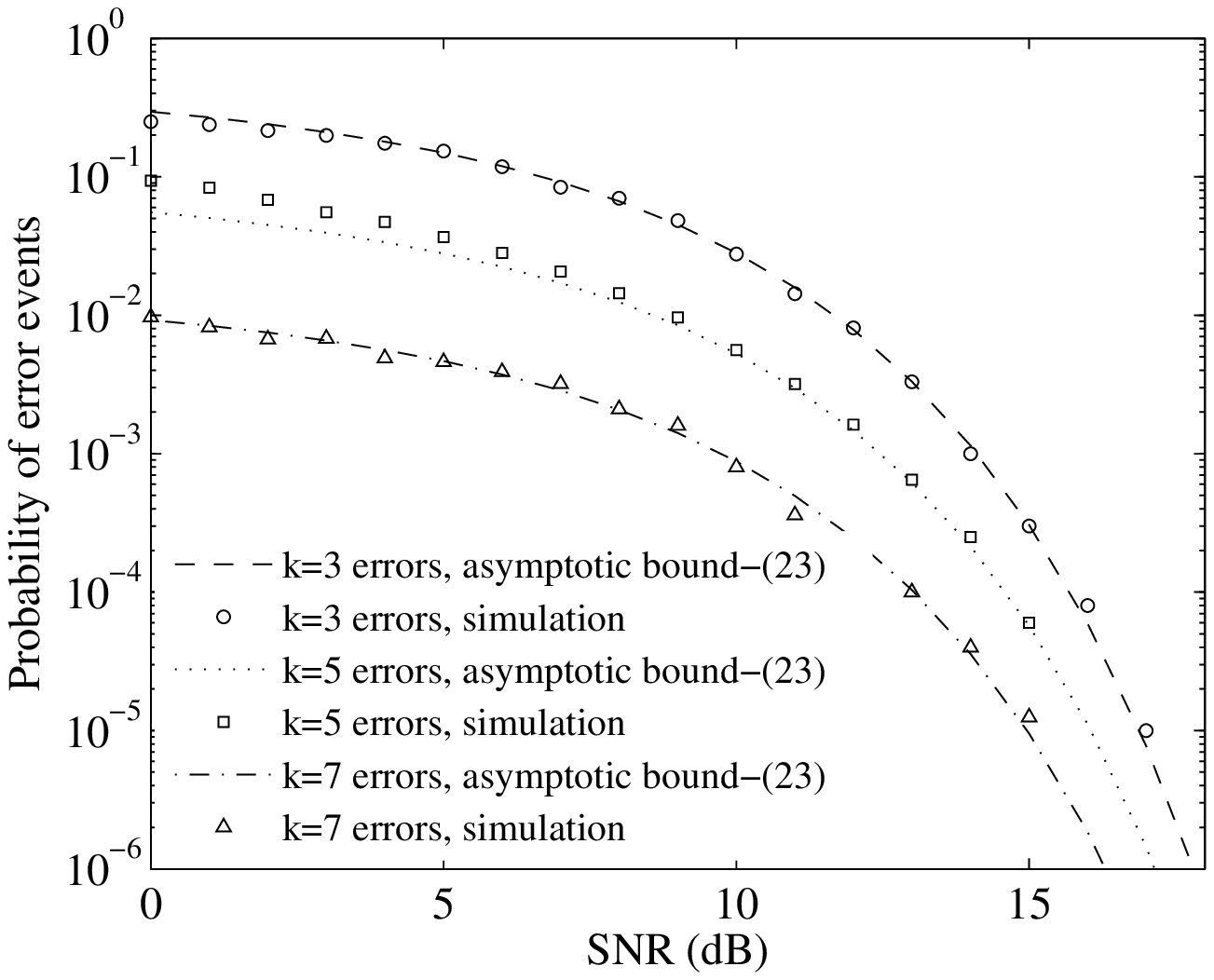}
 \label{fig:su4}}\label{fig:Fig3}
  \caption{Probability of $k$ error events in an $L=10$ user AF MWRN with AWGN.}}
\end{figure}

Figs.~\ref{fig:su3} and~\ref{fig:su4} plot the probability of $k$ error events $P_{i,AF}(k)$ in an $L=10$ user AF MWRN corrupted by AWGN for $k=1,2$ error events and $k=3,5,7$ error events, respectively. The simulation results are plotted for $k=1,2,3,5,7$ and compared with the asymptotic bound in \eqref{51b}. For $k=1,2$ the exact probabilities are also plotted using~\eqref{eq:266} and \eqref{31}, respectively. As highlighted in Remark 8, in an $L$-user AF MWRN, the probability of error events depends on the value of $k$, with the higher order error events being less probable. This is confirmed by the results in Figs.~\ref{fig:su3} and~\ref{fig:su4}. We can see that for medium to high SNR (SNR $> 10$ dB), the asymptotic expression in \eqref{51b} for $k$ error events matches very well with the simulation results. This verifies the accuracy of~\eqref{51b}.

\subsection{Average BER for a user in AWGN DF or AF MWRN}
\begin{figure}
  \includegraphics[width=0.85\textwidth]{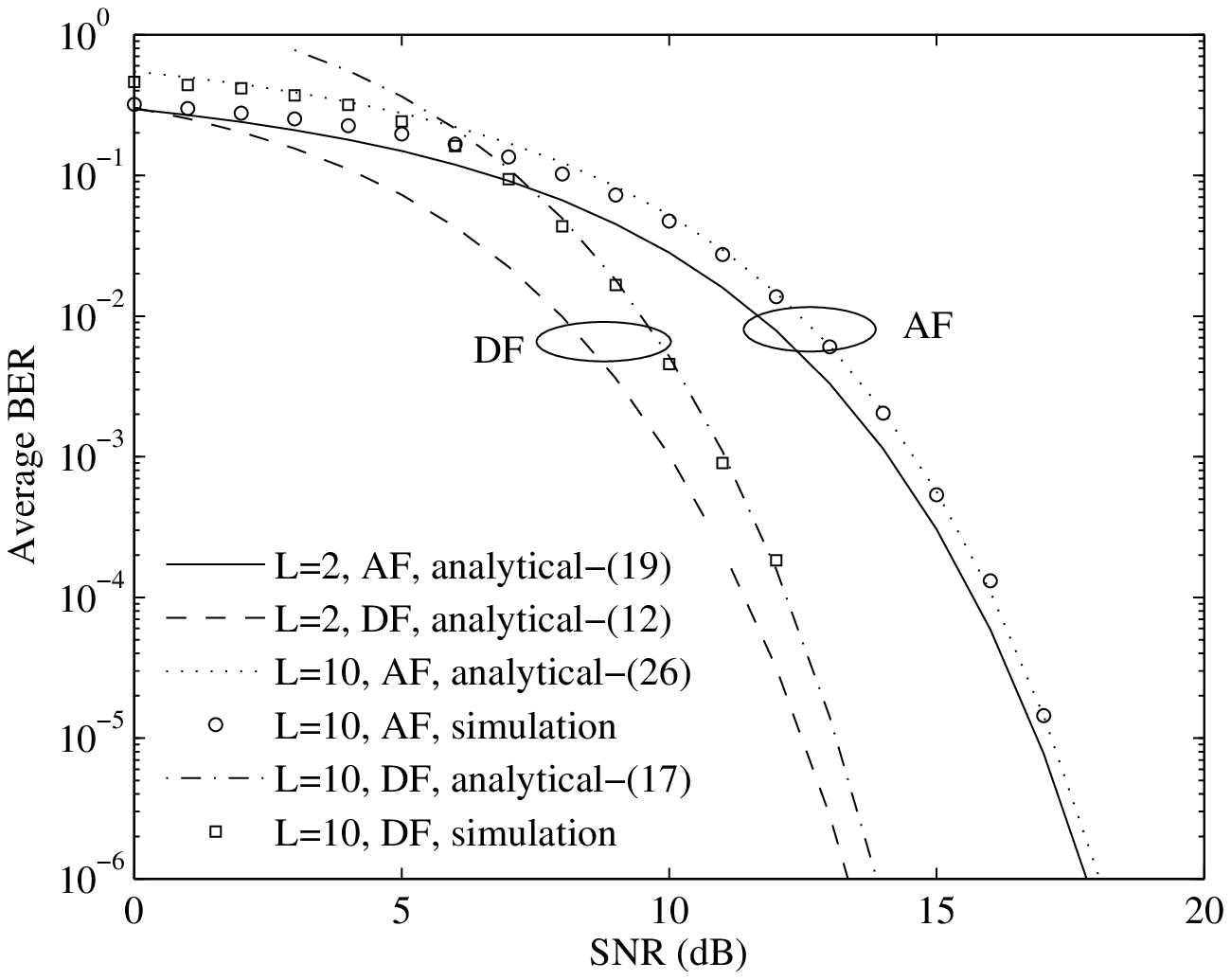}
  \caption{Average BER for a user in an $L=10$ user DF or AF MWRN with AWGN.}
\label{fig:su1}
\end{figure}
\begin{figure}
  \includegraphics[width=0.85\textwidth]{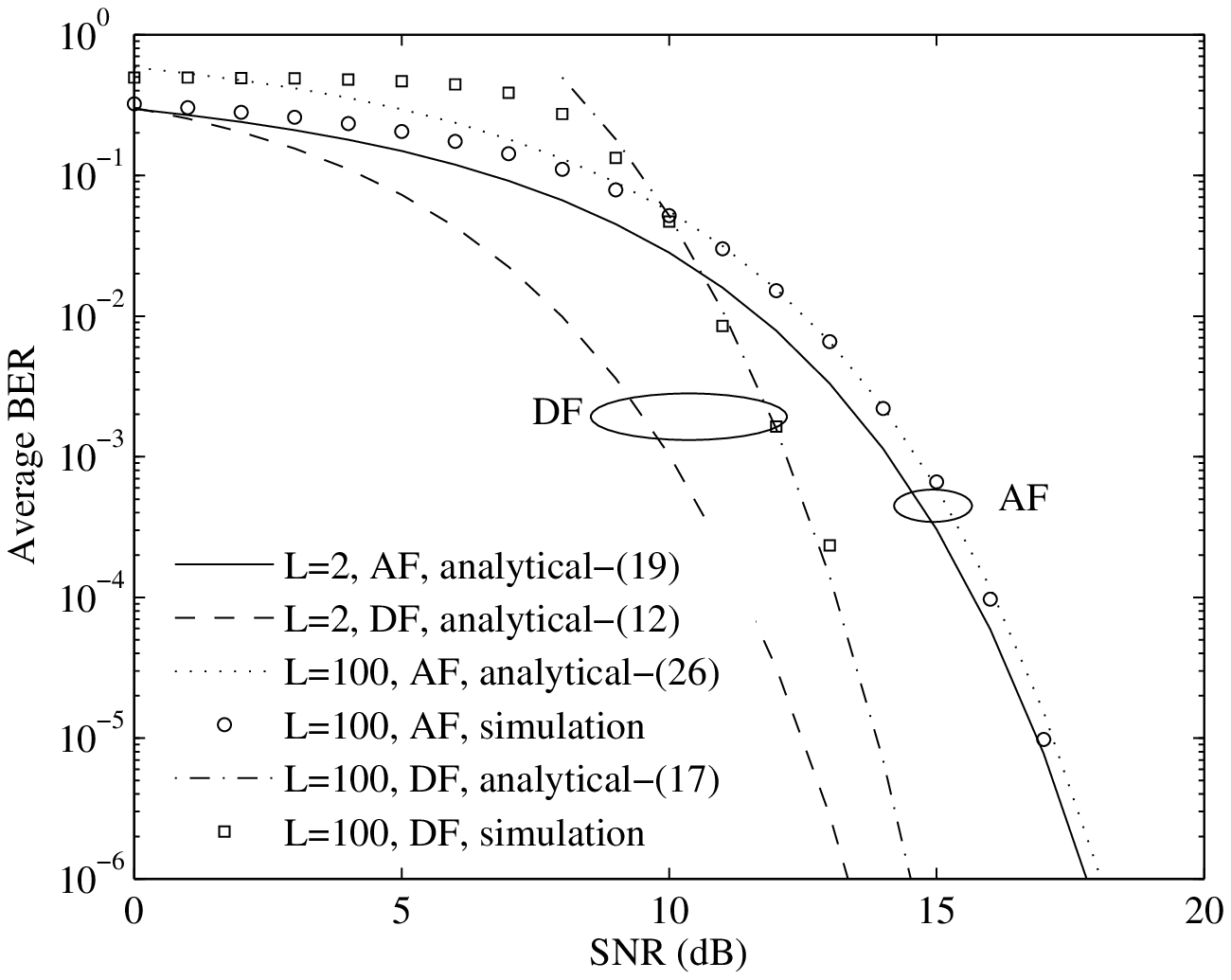}
  \caption{Average BER for a user in an $L=100$ user DF or AF MWRN with AWGN.}
\label{fig:su2}
\end{figure}

Figs.~\ref{fig:su1} and~\ref{fig:su2} plot the average BER for a
user in an AWGN DF or AF MWRN with $L=10$ and $L=100$ users, respectively.
The average BER of DF or AF TWRN, from \eqref{10} or \eqref{15},
respectively, is plotted as a reference. The average BER of DF and
AF MWRN is plotted using \eqref{50a} and \eqref{50b}, respectively.
From the figures, we can see that as the number of users increases
($L=2,10,100$), the average BER increases for both DF or AF MWRN,
which is intuitive. For DF MWRN, \eqref{50a} can predict the average
BER for a user accurately in medium to high SNR (approximately SNR $>
7$ dB for $L=10$ users and SNR $> 10$ dB for $L=100$ users). Also
for AF MWRN, \eqref{50b} can accurately predict the average BER for a user in medium to high SNR (approximately
SNR $> 10$ dB).

Comparing DF and AF MWRNs, we can see that for low SNR, AF MWRN is
slightly better than DF MWRN. However, at medium to high SNRs, DF
MWRN is better than AF MWRN. For TWRN, it can be easily shown that
the high SNR penalty for using AF, compared to DF, is 4.77 dB (see
Appendix A). In MWRN, this high SNR penalty decreases as the number
of users increases, e.g., from Figs.~\ref{fig:su1}
and~\ref{fig:su2}, it is about $4$ dB for $L=10$ users and about
$3.5$ dB for $100$ users. This can be explained using our analysis
as follows. From \eqref{50a} we can see that for DF MWRN the
effective number of error terms in the average BER equation
increases in proportion to the number of users. However, for AF
MWRN, \eqref{50b} shows that the probability of larger number of
error events is very small, hence, the increase in the effective
number of error terms for larger number of users is smaller. This
results in a smaller SNR penalty for AF MWRN when larger number of
users are involved, which agrees with the observations from
Figs.~\ref{fig:su1} and~\ref{fig:su2}.

\subsection{Rayleigh Fading}
\begin{figure}
{\subfigure[$L=10$]{
  \includegraphics[width=0.5\textwidth]{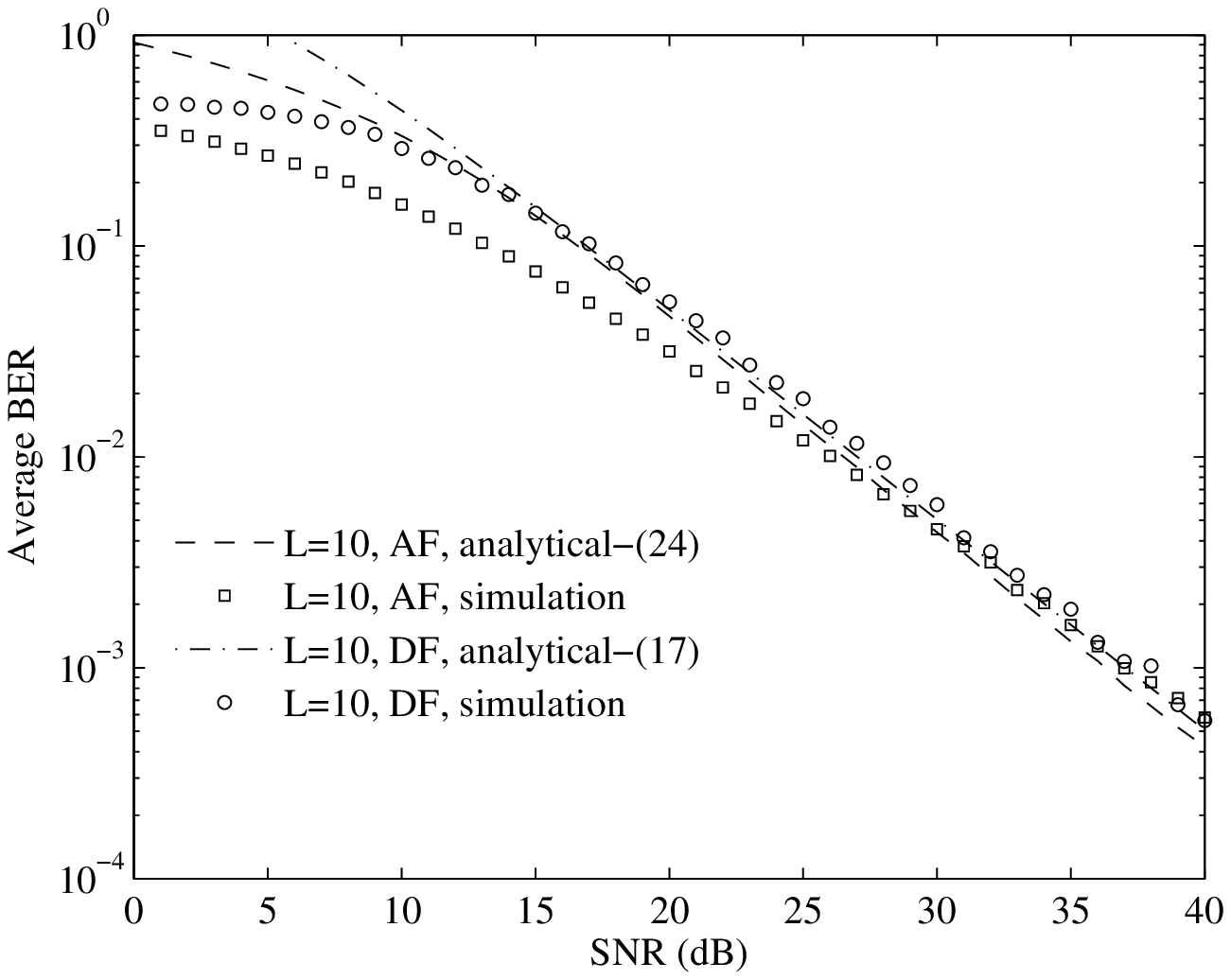}
  \label{fig:su5}}
\subfigure[$L=50$]{
  \includegraphics[width=0.5\textwidth]{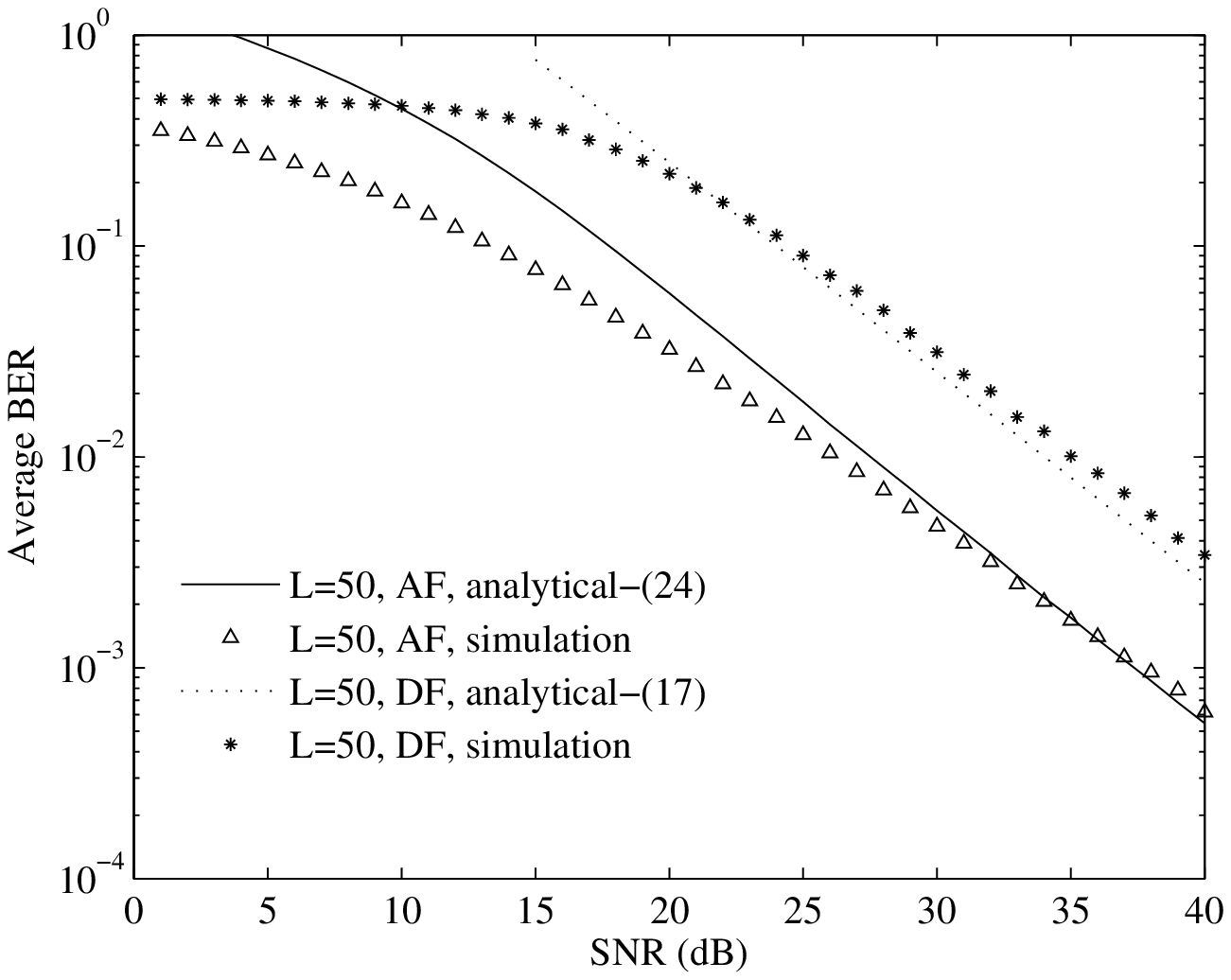}
 \label{fig:su6}}\label{fig:Fig7}
  \caption{Average BER for a user in DF or AF MWRN with Rayleigh fading and $L=10,50$ users.}}
\end{figure}

Figs. \ref{fig:su5} and \ref{fig:su6} plot the average BER for a
user in DF or AF MWRN in Rayleigh fading channels and $L=10$ and $L=50$
users, respectively. The analytical result for DF MWRN is plotted
using \eqref{50a} and \eqref{47} and the analytical result for AF
MWRN is plotted using \eqref{37}, \eqref{51a}, \eqref{53} and
\eqref{60}. We can see that for both DF and AF MWRN the analytical
results are within 1 dB of the simulation results for high
SNR. Comparing the curves for $L=10$ and $L=50$
users, we can see that the average BER for a user in DF MWRN
degrades significantly as the number of users increases. However,
the average BER for a user in AF MWRN is more robust to the increase
in the number of users. As explained before, this is due to the fact
that the probability of larger number of error events in AF is much
smaller compared to DF MWRN.

\section{Conclusions} \label{conclusion}
In this paper, we presented a method for analyzing (i) the probability of $k$ error events and (ii) the
average BER for a user in both DF and AF MWRNs. The method is based on
insights provided by the exact analysis of $k=1$ and $k=2$ error
events, which leads to an accurate asymptotic expression for $k$ error
events in such systems. For both DF and AF MWRN in AWGN channel, the
derived expression can accurately predict the BER of a user in
medium to high SNR. For Rayleigh fading channel, the derived
expressions match with simulations within 1 dB in high SNR.
Using our analysis, we showed that DF MWRN outperforms
AF MWRN in AWGN channels even with a larger number of users, while AF MWRN outperforms DF MWRN in Rayleigh
fading channels even for a much smaller number of users.

\section*{Appendix A\\ Proof of SNR Penalty for using AF}
At high SNR, $\gamma_{r}$ and $\gamma$ can
be approximated to $1$ and $0$, respectively. Substituting these
values in \eqref{10}, we get
$\textrm{erfc}\left(\frac{\gamma_{r}+2}{\sqrt{1/\rho}}\right)\approx0$
and
$\textrm{erf}\left(\frac{\gamma_{r}+2}{\sqrt{1/\rho}}\right)\approx1$.
Based on this, the asymptotic error probability of a DF TWRN can be
given as
\begin{align}\label{55}
    P_{DF}=&\frac{1}{8}\left[\textrm{erfc}\left(\frac{-1}{\sqrt{1/\rho}}\right)\left(1+\textrm{erf}\left(\frac{-1}{\sqrt{1/\rho}}\right)+
    2\textrm{erfc}\left(\frac{1}{\sqrt{1/\rho}}\right)\textrm{erfc}\left(\frac{-1}{\sqrt{1/\rho}}\right)\right)\right.\nonumber\\
    &+\left.\textrm{erfc}\left(\frac{1}{\sqrt{1/\rho}}\right)\left(\textrm{erfc}\left(\frac{-1}{\sqrt{1/\rho}}\right)+
    2\textrm{erf}\left(\frac{1}{\sqrt{\rho}}\right)\textrm{erfc}\left(\frac{1}{\sqrt{\rho}}\right)\right)\right]
\end{align}
Putting $\textrm{erfc}(-x)=2-\textrm{erfc}(x)$ and
$\textrm{erf}(x)=1-\textrm{erfc}(x)$ and after some simplifications,
the above equation can be written as
\begin{align}\label{54}
P_{DF}=&\frac{1}{8}\left[12\textrm{erfc}\left(\frac{1}{\sqrt{1/\rho}}\right)+2\left(\textrm{erfc}\left(\frac{1}{\sqrt{1/\rho}}\right)\right)^{3}-10\left(\textrm{erfc}\left(\frac{1}{\sqrt{1/\rho}}\right)\right)^2+\right.\nonumber\\
&\left.2\left(\textrm{erfc}\left(\frac{1}{\sqrt{1/\rho}}\right)\right)^2\left(\textrm{erf}\left(\frac{1}{\sqrt{1/\rho}}\right)\right)\right]
\end{align}
At high SNR, neglecting the higher order terms, the error
probability of a DF TWRN can be approximated as
\begin{equation}\label{14}
P_{DF,\infty}\approx\textrm{erfc}\left(\sqrt{\rho}\right)
\end{equation}
Similarly, for an AF TWRN, after substituting the value of $\alpha$
in \eqref{15}, the error probability can be approximated at high SNR
as
\begin{equation}\label{16}
P_{AF,\infty}\approx\textrm{erfc}\left(\sqrt{\frac{\rho}{3}}\right)
\end{equation}

Comparing equations \eqref{14} and \eqref{16},
\begin{equation}\label{17}
\textrm{SNR penalty in
AF}=\frac{\rho}{\rho/3}=3=4.77 \textrm{dB}
\end{equation}

%\bibliography{IEEEabrv,ref_journal_9}
\end{document}